\documentstyle[12pt]{article}
\newcommand{\be}{\begin{equation}}
\newcommand{\ee}{\end{equation}}
\newcommand{\ba}{\begin{eqnarray}}
\newcommand{\ea}{\end{eqnarray}}
\begin{document}
%%%%%%%%%%%%%%%%%%%%%%%% FRONTESPIZIO %%%%%%%%%%%%%%%%%%%%%%%%%%%%%%%%%%%%%
\begin{titlepage}
\hbox{\hskip 12cm ROM2F-95/4  \hfil}
\hbox{\hskip 12cm CPTh/RR.355.0395 \hfil}
\hbox{\hskip 12cm March \ 1995 \hfil}
%\end{flushright}
\vskip 1.5cm
\begin{center}

{\Large  \bf    Planar \ Duality \ in \ $SU(2)$ \ WZW \ Models}

\vspace{2cm}

{\large G. Pradisi${}^{a,c}$ , \ A. Sagnotti${}^{a,b}$ \ \ and \ \  Ya.S.
Stanev${}^{a,}$\footnote{I.N.F.N.  Fellow, on
Leave from Institute for Nuclear Research and Nuclear Energy, Bulgarian
Academy of Sciences, BG-1784 Sofia, BULGARIA.}}

\vspace{0.8cm}

${}^{a}${\sl Dipartimento di Fisica\\
Universit\`a di Roma \ ``Tor Vergata'' \\
I.N.F.N.\ - \ Sezione di Roma \ ``Tor Vergata'' \\
Via della Ricerca Scientifica, 1 \ \
00133 \ Roma \ \ ITALY}
\vskip 14pt
${}^{b}${\sl Centre de Physique Th\'eorique \\
Ecole Polytechnique \\
91128 Palaiseau \ \ FRANCE}
\vskip 14pt
${}^{c}${\sl Centro ``Vito Volterra'', Universit\`a di Roma
``Tor Vergata''}
\vspace{1.5cm}
\end{center}

\abstract{
We show how to generalize the $SU(2)$ WZW models to allow for open
and unoriented sectors. The
construction exhibits some novel patterns of Chan-Paton charge
assignments and projected spectra that reflect the underlying
current algebra.}  \vfill
\end{titlepage}
%%%%%%%%%%%%%%%%%%%%%%%%%%%%%%%%%%%%%%%%%%%%%%%%%%%%
\addtolength{\baselineskip}{0.3\baselineskip}

\vskip 24pt
\begin{flushleft}
{\large \bf Introduction}
\end{flushleft}

In a number of previous papers \cite{bs}\cite{bps1}\cite{bps2}\cite{fps} the
idea of associating ``open descendants'' to left-right symmetric
models of oriented closed strings \cite{car} was shown to
determine both the bulk spectra
and the Chan-Paton \cite{cp} charge sectors of new models with
unoriented closed
and open strings.  The link between these two classes of models is a general
feature of Conformal Field Theory,
and indeed open descendants can be constructed \cite{bps1} starting from the
$BPZ$ series \cite{bpz} of minimal models. The close relationship between
the minimal models and the $ADE$ \cite{ciz} models of the $SU(2)$
current algebra \cite{wzw}
suggests to take a closer look at this case as well.  This is particularly
rewarding, since the underlying current algebra introduces a number of novel
features that shed new light on the meaning of the Klein-bottle and M\"obius
projections, while providing a finer test of the ``crosscap''  constraint
of ref. \cite{fps}.
\vskip 24pt
\begin{flushleft}
{\large \bf The $A_3$ Model and its Descendants}
\end{flushleft}

The simplest non-trivial $SU(2)$ model belongs to the $A$ series and
corresponds to $k=2$ \cite{ciz}. It is particularly
instructive,  since both its torus amplitude
\be
T= |\chi_1|^2 \ + \ |\chi_2|^2 \ + \ |\chi_3|^2	\qquad ,
\label{k2t}
\ee
where $\chi_{2I+1}$ denotes the character
corresponding to isospin $I$, and its $S$ matrix
\be
S \, = \, {1 \over 2}
\pmatrix{1&\sqrt{2}&1 \cr \sqrt{2}&0&-\sqrt{2} \cr 1&-\sqrt{2}&1} \quad ,
\label{sk2}
\ee
are mapped into the Ising ones by the identification of
$\chi_1$, $\chi_2$ and $\chi_3$ with the three Ising characters
of identity, spin and energy.
Its open descendants, however, must exhibit a different structure, since
$\chi_2$ has conformal weight $3/16$ while the Ising spin has conformal weight
$1/16$.  Thus, for the Ising model the matrix
\be
P=T^{1/2} \ S \ T^2 \ S \ T^{1/2}
\label{pmat}
\ee
that relates the real bases of characters $\hat{\chi}$ for
direct and transverse M\"obius channels is
\be
P \ = \ \pmatrix{\cos{( \, {\pi \over 8} \, )}&0&\sin{( \, {\pi \over 8} \, )}
\cr
0&1&0 \cr
\sin{( \, {\pi \over 8} \, )}&0&- \, \cos{( \, {\pi \over 8} \, )} \cr
} \qquad  ,
\label{isingp}
\ee
while for the $A_3$ model the $P$ matrix is
\be
P \ = \ \pmatrix{\sin{( \, {\pi \over 8} \, )}&0&\cos{( \, {\pi \over 8} \, )}
\cr
0&1&0 \cr
\cos{( \, {\pi \over 8} \, )}&0&- \, \sin{( \, {\pi \over 8} \, )} \cr
} \qquad  .
\ee

This state of affairs is reminiscent of the behavior of some
models discussed in ref. \cite{bs}, where off-diagonal $P$ matrices result in
the appearance of ``complex'' Chan-Paton charges.  Indeed, this model can
accommodate a real charge and a pair of complex
charges, since the annulus amplitude
\be
A \ = \ \biggl(  {{n_2^2}\over{2}} + m \bar{m} \biggr) \ \chi_1 \ + \
n_2 \bigl( m + \bar{m} \bigr) \ \chi_2 \ + \
{{n_2^2 + m^2 +\bar{m}^2} \over 2} \ \chi_3
\label{acompl}
\ee
and the M\"obius amplitude
\be
M \ = \ \pm \ \biggl[ \ {{n_2}\over{2}} \ \hat{\chi}_1 \ + \
{{n_2 + m +\bar{m}} \over 2} \ \hat{\chi}_3 \ \biggr]
\label{mcompl}
\ee
are consistent both in the direct and in the transverse channel if the
Klein bottle {\it symmetrizes} all three Verma modules
corresponding to $\chi_1$, $\chi_2$ and $\chi_3$.

For the sake of comparison,
in the Ising model the same choice of Klein-bottle
projection leads to \cite{bps1}
\be
A \ = \ \biggl(  {{n_0^2 + n_{1/2}^2 + n_{1/16}^2}\over{2}} \biggr) \chi_0 \ +
\
n_{1/16} ( n_0 + n_{1/2} ) \chi_{1/16} \ + \ \biggl({{n_{1/16}^2} \over 2} +
n_0 n_{1/2} \biggr) \chi_{1/2}  \label{aisingr}
\ee
and
\be
M \ = \ \pm \ \biggl[ \ {{n_0 + n_{1/16} + n_{1/2}}\over{2}} \
\hat{\chi}_0 \ + \ {{n_{1/16}} \over 2} \ \hat{\chi}_{1/2}
\ \biggr]
\qquad . \label{misingr}
\ee

Since the charge assignments of eqs. (\ref{acompl}) and (\ref{mcompl})
do not follow
the pattern suggested by Cardy's analysis \cite{cardy} of the Verlinde
formula \cite{ver}, one may wonder whether a different Klein bottle
projection could result in a model with all real charges.  This actually
corresponds to the only other choice of closed spectrum
compatible with the positivity of the vacuum Klein-bottle channel,
\be
K \ = \ {1 \over 2} \biggl( \chi_1 - \chi_2 + \chi_3 \biggr) \qquad ,
\label{kreal}
\ee
whereby all the integer-isospin states are symmetrized while
all the half-odd-integer-isospin ones are antisymmetrized.  Indeed, in the
resulting model,
\be
A \ = \ \biggl(  {{n_1^2 + n_2^2 + n_3^2}\over{2}} \biggr) \chi_1 \ +
\
n_2 ( n_1 + n_3 ) \chi_2 \ + \ \biggl({{n_2^2} \over 2} +
n_1 n_3 \biggr) \chi_3  \label{areal}
\ee
and
\be
M \ = \ \pm \ \biggl[ \ {{n_1 - n_2 + n_3}\over{2}} \ \hat{\chi}_1 \ +
\ {{n_2} \over 2} \ \hat{\chi}_3 \ \biggr]
\label{mreal}
\ee
are consistent choices both in the direct and in the transverse
channel.  The annulus amplitude now reflects the
fusion rules, but the M\"obius amplitude involves
some alternating signs whose origin is quite interesting and, as we
shall  see, bears a close relationship to the underlying current
algebra.

One is now encouraged to repeat the same exercise for the Ising model, since
after all the two $A_3$ models differ in their Klein-bottle projection in a
way that can be traced to $Z_2$, the center of $SU(2)$, that
distinguishes between integer and half-odd-integer
isospin representations and manifests itself in a corresponding
automorphism of the fusion algebra.
A similar $Z_2$ symmetry is present in the Ising model as well, and indeed
one can construct a new class of open descendants starting from
\be
K = {1 \over 2} \biggl( \chi_0 - \chi_{1/16} +
\chi_{1/2} \biggr) \label{kisingc} \qquad ,
\ee
again the only other projection of the closed
spectrum compatible with the positivity of the vacuum Klein-bottle channel.
The resulting model involves one real charge and a pair of complex charges,
and
\be
A \ = \ \biggl(  {{n_{1/16}^2}\over{2}} + m \bar{m} \biggr) \ \chi_0 \ + \
n_{1/16} \bigl(  m + \bar{m} \bigr) \ \chi_{1/16} \ + \
{{n_{1/16}^2 + m^2 +\bar{m}^2} \over 2} \ \chi_{1/2}
\label{aisingc}
\ee
and
\be
M \ = \pm \ \biggl[{{n_{1/16}}\over{2}} \ \hat{\chi}_0 \ + \
{{ m +\bar{m} - n_{1/16}} \over 2}  \ \hat{\chi}_{1/2} \ \biggr]
\label{misingc}
\ee
are a consistent choice both in the direct and in the transverse channel.  The
models in ref. \cite{bps1}  may thus be extended to include (infinitely many)
others with a different Klein-bottle projection and with pairs of real
charges replaced by complex ones \cite{pss}.

These results deserve some discussion, since the ``crosscap'' constraint
of ref. \cite{fps} singles out the usual Klein-bottle projection for the
Ising model, with {\it all} three sectors symmetrized.  Whereas this would
appear to exclude the model of eqs. (\ref{kisingc}), (\ref{aisingc})
and (\ref{misingc}), in some cases the constraint may actually be
relaxed to some extent.  This is possible whenever the fusion algebra
allows some of the two-point
functions in front of a crosscap $< \phi_{{h_1},\bar{h}_1} \
\phi_{{h_2},\bar{h}_2} >_c$ to behave as $Z_2$ sections \cite{pss}.
Thus, including a
sign $\epsilon$ depending on the (symmetric or antisymmetric) nature of
one of the two fields, say $\phi_{{h_1},\bar{h}_1}$, in the projected closed
spectrum, the ``crosscap'' constraint reads
\be
\epsilon_{(1,\bar{1})} \ (-1)^{h_1 - \bar{h}_1 + h_2 - \bar{h}_2} \ C_{12k} \
C_{\bar{1}
\bar{2} k} \ \Gamma_k \ = \ \sum_p \ C_{\bar{1}2p} \ C_{1 \bar{2} p} \
\Gamma_p \ F_{pk}(1,2,\bar{1},\bar{2}) \qquad ,
\label{crosscap}
\ee
where $C_{ijk}$ are the chiral
structure constants, $\Gamma_{k}$ are the crosscap one-point coefficients and
$F_{pk}$ is the duality matrix that relates the $s$-channel conformal
blocks to
the $u$-channel ones.  We would like to stress that eq. (\ref{crosscap})
contains more information than the Klein-bottle amplitude,
that involves only the squared one-point coefficients.
In the Ising model with complex charges this generalized
constraint is fulfilled by {\it all} two-point functions, with a positive sign
if
$\phi_{{h_1},\bar{h}_1}$ is the identity or the energy, both symmetrized by the
Klein-bottle projection, but with a {\it negative} sign if
$\phi_{{h_1},\bar{h}_1}$ is the spin. A more detailed discussion
will be presented in ref. \cite{pss}, where eq. (\ref{crosscap})
will be shown to
determine the structure of some exotic open descendants
corresponding to the $E_7$ model and to the $D_{odd}$ series. Still, we should
anticipate that in WZW models for $\epsilon=1$ integer-isospin sectors are
symmetrized and half-odd-integer ones are antisymmetrized, while for the
other possible choice,
$\epsilon=(-1)^{2I}$, all sectors are symmetrized.  The results of ref.
\cite{rst},
where the duality matrices for WZW models are constructed explicitly, reveal
the
general occurrence of this phenomenon, since the direct Klein-bottle projection
involves $SU(2)$ singlets, that originate from (anti)symmetric combinations for
(half)integer spins.

Similar remarks apply to the relation between the ``twist'' properties of
the open states in the Ising and $A_3$ WZW models just discussed.
Indeed, one may wonder how the new model of eqs. (\ref{aisingc})
and (\ref{misingc}) can
be compatible with the duality of disk amplitudes.  The puzzle in this respect
may be stated as follows. Once the behavior of the prefactor is disposed of
by a suitable prescription \cite{bps1}, the conformal blocks
should apparently determine the duality properties of all amplitudes.
Thus, for instance, it
is not obvious how the relative ``twist'' of energy and identity states with
two
$n_{1/16}$ charges, opposite in the model of
eqs. (\ref{aisingc}) and (\ref{misingc}), may be identical in the model of
eqs. (\ref{aisingr}) and (\ref{misingr}). A closer inspection is
instructive, since it reveals how the non-trivial
action of the ``twist'' on complex Chan-Paton charges actually
ensures the consistency of both settings.
\vskip 30pt
\input epsf \centerline{ \epsfbox{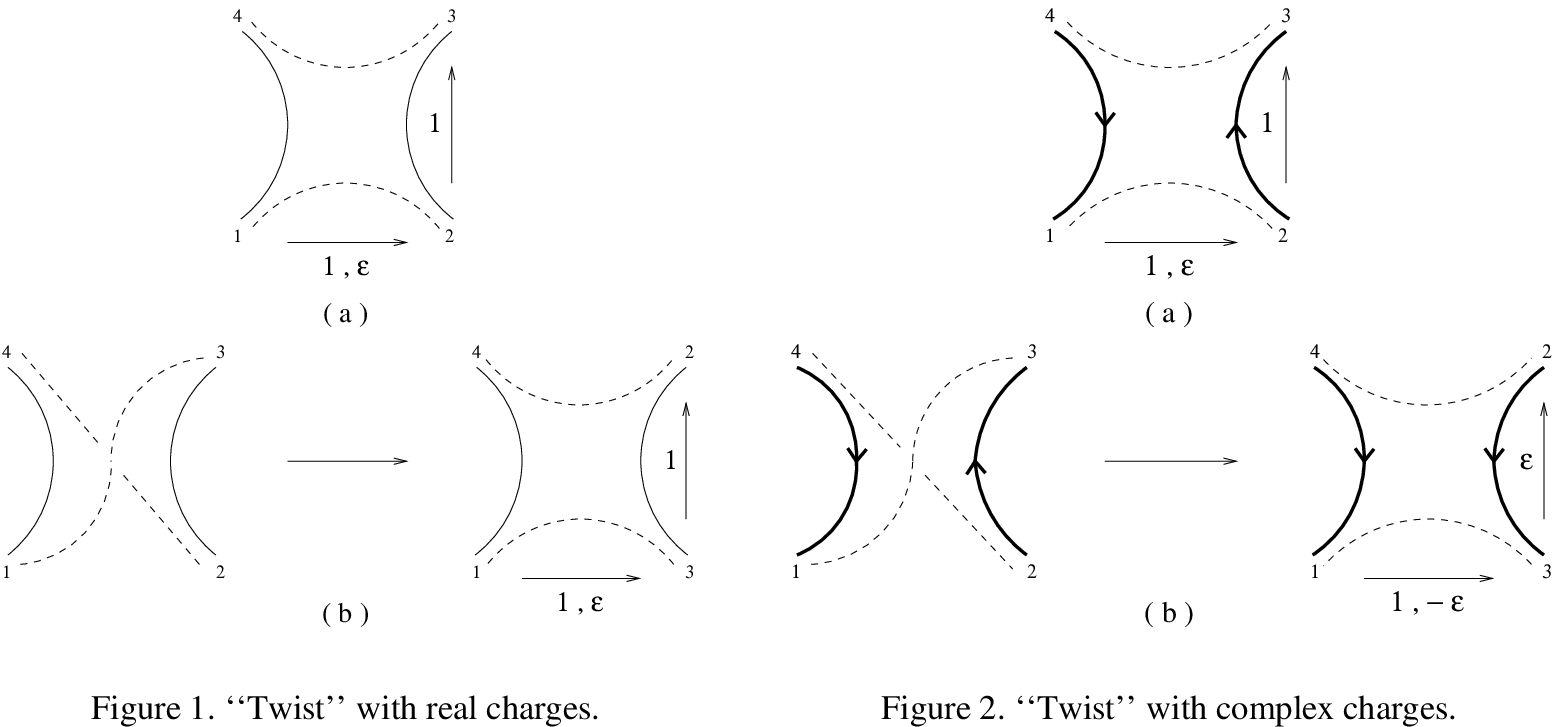}}
\vskip 20pt

Referring to the four-spin amplitude of fig. 1, where dashed
lines denote $n_{1/16}$ charges and continuous lines denote $n_0$
charges, let us observe that only the identity flows in the $s$ channel
(fig. 1a), while both the identity and the energy flow in the $u$
channel.  Aside from a prefactor, the amplitude of fig. 1a is \be A_a
\ = \ Tr({\Lambda_1}^T \ {\Lambda_2} \ {\Lambda_3}^T \ {\Lambda_4})
\   \sqrt{1 + \sqrt{1-x}} \qquad , \label{aas} \ee where the limit of
small values for the cross-ratio  $x = {\small{{z_{12} z_{34}} \over
{z_{13} z_{24}}}}$ exhibits the $s$ channel, while the transformation
$x \rightarrow (1-x)$ exposes  the $u$ channel.  The duality
transformations of the conformal blocks amount in this case to a
familiar identity and, dropping again a prefactor, $A_a$ becomes
\be
A_a \ = \  Tr(  {\Lambda_4} \ {\Lambda_1}^T \ {\Lambda_2} \
{\Lambda_3}^T ) \
\biggl(\sqrt{1 +  \sqrt{1-x}} + \sqrt{1 -
\sqrt{1-x}} \ \biggr) \qquad .
\label{aau}
\ee

The corresponding ``twisted'' $u$-channel contribution of fig. 1b
unfolds into an amplitude where the identity flows in the $t$ channel while,
again, both the identity and the energy flow in the $u$ channel, with the {\it
same} sign.
Indeed, following ref. \cite{schwarz}, the ``twisted'' contribution may be
exposed by performing the transformation $x \rightarrow x/(1+x)$
in an amplitude obtained from eq. (\ref{aau}) by the
interchange of the external legs $2$ and $3$.  This operation may be
regarded as the
the braiding $B_1 : x \rightarrow e^{i \pi} \ x/(1-x)$ followed by the
reflection $x \rightarrow e^{- i \pi} \ x$.  The
``twisted'' contribution is then
\be
A_a^{(tw)} \ = \ Tr(  {\Lambda_4} \ {\Lambda_1}^T \ {\Lambda_3} \
{\Lambda_2}^T ) \ \biggl(\sqrt{1 +  \sqrt{1+x}} + \sqrt{\sqrt{1+x}-1} \
\biggr) \qquad ,  \label{aat}
\ee
and thus {\it for all levels} the identity and energy states with
a pair of $n_{1/16}$ charges have identical ``twist'' properties,
consistently with the M\"obius amplitude of eq. (\ref{misingr}).

The four-spin amplitude for the Ising model with complex charges
is displayed in fig. 2.  As before,
dashed lines denote $n_{1/16}$ charges, but now the remaining
lines carry arrows associated to the ``complex'' charges $m$ and
$\bar{m}$, and eq. (\ref{aau}) is replaced by
\be
A_b \ = \ Tr({M_1}^{\dagger} \ {M_2} \ {M_3}^{\dagger} \
{M_4}) \ \biggl(\sqrt{1 +
\sqrt{1-x}} + \sqrt{1 -
\sqrt{1-x}} \ \biggr) \qquad .
\label{acu}
\ee
However, unfolding  the ``twisted'' diagram now inverts the relative
orientation  of the arrows (fig. 2b) and, consistently with the
partition function, the resulting amplitude
\be
A_b^{(tw)} \ = \ Tr({M_1}^{\dagger} \ {M_3}^{\ast} \ {M_2}^{T} \ {M_4}) \
\sqrt{1 - \sqrt{1-x}}
\label{acs}
\ee
propagates the energy, not the identity, in the
$t$ channel.  Therefore, the ``twisted'' $u$-channel contribution
to $A_b$,
\be
A_b^{(tw)} \ = \ Tr({M_1}^{\dagger} \ {M_3}^{\ast} \ {M_2}^{T} \ {M_4}) \
\biggl(\sqrt{1 + \sqrt{1+x}} - \sqrt{\sqrt{1+x}-1} \ \biggr) \qquad  ,
\label{act}
\ee
now involves energy and identity with {\it opposite} signs.

Returning to the $A_3$ WZW models, we would like to relate the additional
signs in the M\"obius amplitude to the behavior of the $SU(2)$
current algebra blocks.  We confine our attention to the $A_3$
model with real charges of eqs. (\ref{areal}) and (\ref{mreal}), but
similar considerations apply to the model of eqs. (\ref{acompl})
and (\ref{mcompl}). Leaving aside a
prefactor, and referring again to fig. 1a, where now dashed
lines denote $n_2$ charges and continuous lines denote $n_1$
charges, the $u$-channel amplitude for four isospin-1/2
states may be written
\be
A_a \ = \ Tr({\Lambda_1}^T \ {\Lambda_2} \ {\Lambda_3}^T \ {\Lambda_4})
\ \biggl(
S_1( x, \xi ) - S_0( x , \xi ) \biggr) \qquad , \label{ask2}
\ee
where
\ba
S_0( x, \xi ) \  & = & \ s_0( \xi ) \ \biggl[ \  (1 - x)^{1/2} \ \sqrt{1  +
\sqrt{1-x}} \  +  \ {1 \over 2} x^{1/2} \ \sqrt{1 - \sqrt{1-x}} \ \biggr] \ -
\nonumber \\
& & {s_1( \xi )} \ x^{1/2} \ \sqrt{1 - \sqrt{1-x}} \qquad  \qquad ,
\label{S0} \\
S_1( x, \xi ) \ & = & \ s_0( \xi ) \ \biggl[ \  (1 - x)^{1/2}
\ \sqrt{1  -  \sqrt{1-x}} \  -  \ {1 \over 2} x^{1/2} \ \sqrt{1 +
\sqrt{1-x}} \ \biggr] \ + \nonumber \\
& & {s_1( \xi )} \ x^{1/2} \ \sqrt{1 +
\sqrt{1-x}} \qquad  \qquad  , \label{S1}
\ea
and
\be
s_0( \xi ) \ = \ \xi \qquad {\rm and} \qquad s_1( \xi ) \ = \ 1 - {1 \over 2}
\ \xi
\qquad ,
\label{s01}
\ee
with $x$ the cross ratio defined above and $\xi$ a corresponding cross-ratio
of new auxiliary variables \cite{bt} present in the chiral vertex operators of
the external states.

The auxiliary variables are particularly convenient, since they
lead to compact expressions for the chiral vertex
operators in terms of polynomials, where the powers select the different
isospin projections. Indeed, if $\varphi_I^{I_3}(z)$ is a vertex
operator of isospin
$I$ and third isospin projection $I_3$, defining
\be
\varphi_I(z,\zeta) \ = \ \sum_{m=-I}^I {\zeta^{I+m} \over (I+m)!}
\varphi_I^m(z) \qquad ,
\label{compzeta}
\ee
the $SU(2)$ invariants entering the 4-point amplitude (\ref{ask2}) are
polynomials in $\zeta_{ij}=\zeta_i - \zeta_j$ or,
apart
form a common prefactor (in our case equal to $\zeta_{13}\zeta_{24}$),
polynomials in  $\xi = {\zeta_{12}\zeta_{34} \over
\zeta_{13}\zeta_{24}}$. The expressions $s_I(\xi)$ in eq. (\ref{s01})
thus correspond to fields with fixed isospin $I$ flowing in the
s channel, and one may deal with the
``twist'' properties of all components at the same time. Moreover, since the
$\zeta$ variables are {\it inert} under the reflection, the
transformation $x \rightarrow x/(1+x)$ should be accompanied by
$\xi \rightarrow {- \xi}/(1-{\xi})$.  Apart from a
common overall factor disposed of by
a suitable prescription, this alters $s_0$ and $s_1$ according to
\be
s_0 \rightarrow - s_0 \qquad {\rm and} \qquad s_1 \rightarrow s_1 \qquad ,
\label{extratw}
\ee
and the additional sign is precisely responsible for the different ``twist''
properties of the Ising and $A_3$ models.
Similar considerations apply to the model
with complex Chan-Paton charges where, as in the Ising model,
unfolding the ``twisted''
amplitude leads to a different $t$ channel contribution. In the duality
matrices for the blocks these differences result in relative
signs between the elements of the two lines.  Thus, for the Ising model
the matrix $F$ of eq. (\ref{crosscap}) for $(0,1/2)$ is
\be
F \ = \ {1 \over {\sqrt{2}}} \
\pmatrix{1&{ \ \ 1}\cr
1&{- 1}\cr } \qquad  ,
\label{isingf}
\ee
while for the $A_3$ WZW model the matrix for $(S_0,S_1)$ is
\be
F \ =  \ {1 \over {\sqrt{2}}} \
\pmatrix{{-1}&1\cr
{\ \ 1}&1\cr } \qquad  .
\label{k2f}
\ee
These observations underlie the constructions presented in the
next Sections.
\vskip 24pt
\begin{flushleft}
{\large \bf The $A$ Series}
\end{flushleft}

The previous results may be extended to the whole $A$ series of
modular invariants.  As compared to ref. \cite{bps1},
these amplitudes contain fewer terms but involve a number of additional
subtleties that have essentially
emerged in the analysis of the $A_3$ case.

For a generic $A$ model
corresponding to level $k$ the $S$ matrix is
\be
S_{ab} \ = \ \sqrt{2 \over {k+2}} \ \sin \biggl( {{\pi a b} \over
{k + 2}} \biggr)
\qquad ,
\label{smatrix}
\ee
while the $P$ matrix is
\be
P_{ab} \ = \ {2 \over {\sqrt{k+2}}} \ \sin \biggl( {{\pi a b} \over
{2 (k + 2)}} \biggr)
( E_k E_{a+b} + O_k O_{a+b} )
\qquad ,
\label{pmatrix}
\ee
with $E$ and $O$ even and odd projectors respectively.
Starting from the torus amplitude
\be
T \ = \ \sum_{a=1}^{k+1} |\chi_a|^2 \qquad ,
\label{torusa}
\ee
the Klein-bottle projection leading to all real charges is
\be
K \ = \ {1 \over 2} \ \sum_{a=1}^{k+1} (-1)^{(a-1)} \chi_a
\label{kleinar}
\ee
where, again, the label $a$ corresponds to $2I+1$, with $I$ the
isospin.  As for the $A-A$ series of minimal models, the
direct-channel annulus amplitude
\be
A \ = \ {1 \over 2} \ \sum_{a,b,c} \ N_{ab}^{c} \ n^{a} \
n^{b} \ \chi_{c}	\qquad ,
\label{annulusar}
\ee
is determined by the fusion-rule coefficients according to the ansatz of
ref. \cite{bs}. Eqs. (\ref{kleinar}) and (\ref{annulusar})
then fix the transverse
M\"obius amplitude, and once the resulting projection
is expressed in terms of the fusion-rule coefficients, one obtains
a rather pleasing expression, namely
\be
M \ = \ \pm {1 \over 2} \ \sum_{a,b} \  (-1)^{b-1} \ (-1)^{{a-1} \over
2} \ N_{bb}^{a} \ n^{b} \ \hat{\chi}_{a} \qquad , \label{moebiusar}
\ee
where $N_{bb}^{a}$ lets only integer-isospin
states flow in the M\"obius strip.  The two phase factors are
very interesting, and could both be anticipated in view of the discussion
of the $A_3$ models. The first phase, $(-1)^{{a-1} \over
2}$, accounts for the different behavior of integer and
half-odd-integer isospin states in the presence of a crosscap,
and is the ``square root'' of a similar phase in the Klein-bottle
amplitude of eq. (\ref{kleinar}).
The second phase, $(-1)^{b-1}$, is even more interesting, since it
distinguishes among the
types of Chan-Paton charges according to the isospin of the
corresponding characters.  Moreover,
it is properly the square of the previous one, since in flowing along
the boundary of the M\"obius strips the charges effect two complete turns
about the crosscap \cite{schwarz}.

On the other hand, starting from a totally symmetric closed spectrum, so that
\be
K \ = \ {1 \over 2} \ \sum_{a=1}^{k+1} \chi_a \qquad ,
\label{kleinac}
\ee
complex charges appear.  If $k$
is even, the model contains an odd number of characters in the
annulus amplitude and an odd number of charges.  The charge
corresponding to the
middle character $\chi_{(k+2)/2}$ stays real, while the charges corresponding
to $\chi_a$ and $\chi_{k+2-a}$ form complex pairs.  On the other hand, if $k$
is odd {\it all} charges form $(k+1)/2$ complex pairs.  In both cases,
all signs disappear from the M\"obius
projection, and the resulting open spectrum is described by
\be
A \ = \ {1 \over 2} \sum_{a,b,c} \ N_{ab}^{c} \ n^{a} \ n^{b} \
\chi_{k+2-c}
\label{annulusac}
\ee
and
\be
M \ = \ \pm {1 \over 2} \sum_{a,b} \ N_{bb}^{a} \ n^{b} \
\hat{\chi}_{k+2-a} \qquad .
\label{moebiusac}
\ee
As usual whenever complex charges are present \cite{bs},
the identifications $n_{k+2-a} = {\bar{n}}_a$, implicit in eqs.
(\ref{annulusac}) and (\ref{moebiusac}), ensure the positivity
of the annulus vacuum channel.  These conditions endow the corresponding
boundaries with an orientation, in the spirit of the mechanism
displayed in fig. 2.
\vskip 24pt
\begin{flushleft}
{\large \bf Other Models}
\end{flushleft}

In extending the construction to the other classes of WZW models, one has
two distinct options for the projection of the closed spectrum whenever
fields of half-integer isospin are present,
to wit in the $E_7$ and $D_{odd}$ cases. Though consistent
with the crosscap
constraint and with the factorization of disk amplitudes, the open
sectors of these models are {\it not} directly based on
the fusion algebra, and the independent charge sectors are
fewer than one would naively expect \cite{pss}.

The other descendants correspond to the $E_{even}$ and $D_{even}$
models, all of which have an extended symmetry.
They follow the pattern dictated by the fusion rules for the
characters of the extended algebra, and can all
be constructed
systematically once one resolves the ambiguity by carefully extending $S$
and  $P$ so that in all cases
\be
( S )^2 \ = \ ( S T )^3 \ = \ ( P )^2 \ = \ C \qquad , \label{sandp}
\ee
where $C$ is the charge-conjugation matrix.
This introduces occasional factors of two in the fusion rules, and
thus the $D_{even}$ models with $k=8p$ have a single set of descendants
with corresponding
factors of two (for $p \geq 2$) in the direct annulus
and M\"obius amplitudes, that reflect the
occurrence of more than one three point function for some sets of fields. In
the $D_{even}$ models with $k=8p+4$ and $p \geq 1$ this amusing new feature is
accompanied by the more familiar
occurrence of a complex pair of Chan-Paton charges \cite{bs} associated to
their
two mutually conjugate characters, that we shall denote
$\chi_{k/2 +1}$ and $\tilde{\chi}_{k/2 +1}$.

In order to exhibit the additional factors of two that the extended
algebra introduces in the annulus
amplitude, it is sufficient to display the annulus and M\"obius amplitudes of
the $k=16$ $D_{even}$ model, while keeping only the two charges
corresponding to the generalized
characters $\chi_c = \chi_5 + \chi_{13}$
and $\chi_d = \chi_7 + \chi_{11}$. Then,
letting $\chi_a = \chi_1 + \chi_{17}$, $\chi_b = \chi_3 + \chi_{15}$,
denoting by $\chi_e$ and $\chi_{\tilde{e}}$ the two ``resolved''
characters, and choosing for definiteness an overall positive sign for the
M\"obius amplitude,
\ba
A  & = &  {{n_c^2 + n_d^2} \over 2}  \ \chi_a \  +  \
\biggl( {{n_c^2 + n_d^2} \over 2} \ + \ n_c n_d \biggr) \
( \chi_b  \ + \ \chi_e
\ + \ \chi_{\tilde{e}} ) \ + \nonumber \\
& & \biggl( {{n_c^2 + 2 n_d^2} \over 2} \ + \ n_c n_d \biggr) \ \chi_c \ + \
\biggl( {{n_c^2 + 2 n_d^2} \over 2} \ + \ 2 n_c n_d \biggr) \ \chi_d
\label{ad16}
\ea
and
\ba
M \ = \  {{n_c + n_d} \over 2} \ ( \hat{\chi}_a - \hat{\chi}_b +
\hat{\chi}_e + \hat{\chi}_{\tilde{e}} )
\ + \ {{n_c + 2 n_d} \over 2} \ \hat{\chi}_c \ - \ {{n_c} \over 2} \
\hat{\chi}_d
\qquad .
\label{md16}
\ea

This example
exibits rather neatly three types of unconventional
Chan-Paton multiplicities.  The first presents itself in the open states
described by $\chi_c$, where factors of two occur both in the annulus and
in the M\"obius amplitude for the charges of type $d$.
There are thus {\it two families} of such states.
The other two present themselves in
the open states corresponding to $\chi_d$, where the annulus amplitude
contains factors of two both for $n_d^2$ and for $n_c n_d$.
Since the M\"obius amplitude does not contain $n_d$, there are {\it two
sectors} of states with a pair of charges of type $d$, described by
symmetric
and antisymmetric matrices respectively, as well as {\it two
sectors} of states with
a pair of distinct charges, of types $c$ and $d$.  These multiple sets
of states reflect the occurrence in these models of multiple three point
functions, a consequence of the extended symmetry.

We would like to conclude by displaying an alternative
way to present our results directly in terms of the $S$ and $P$ matrices,
that applies to all models discussed in this paper. Denoting
by $\tilde{K}$, $\tilde{A}$ and $\tilde{M}$ the transverse-channel amplitudes,
the expressions
\ba
K \ & = & \ {1 \over 2} \ \sum_{a,b} \ \chi_a \
{{ (P_{1b})^2 \ {S^\dagger}_{ab}} \over
{S_{1b}}} \qquad , \label{kclass} \\
\tilde{K} \ & = & \ {1 \over 2} \sum_{a} \ \chi_a \ {\biggl( {{P_{1a}} \over
{\sqrt{S_{1a}}}} \biggr)}^2 \qquad , \label{ktclass} \\
A \ & = & \ {1 \over 2} \ \sum_{a,b,c} \ \chi_a \ n^b n^c
\biggl( \sum_{d} {{{S^{\dagger}}_{ad} \ S_{bd} \ S_{cd}}
\over {S_{1d}}} \biggr) \qquad , \label{aclass} \\
\tilde{A} \ & = & \ {1 \over 2} \ \sum_{a} \ \chi_a \
{\biggl( \sum_{b} {{S_{ab} \ n^b} \over
{\sqrt{S_{1a}}}} \biggr)}^2 \qquad , \label{atclass} \\
M \ & = & \ \pm \ {1 \over 2} \ \sum_{a,b} \ \hat{\chi}_a \ n^b
\biggl( \sum_{d} {{P_{1d}
\ S_{bd} \ {P^\dagger}_{ad}}
\over {S_{1d}}} \biggr) \qquad , \label{mclass} \\
\tilde{M} \ & = & \ \pm \ {1 \over 2} \ \sum_{a,b} \
\hat{\chi}_a \ \biggl({{P_{1a} \ S_{ab} \ n^b}
\over {S_{1a}}} \biggr) \qquad , \label{mtclass}
\ea
(where in the $E_{even}$ and $D_{even}$ cases $S$ and $P$ are the modular
matrices of the extended algebra),
describe consistent open and unoriented spectra.  Moreover,
they relate the crosscap
coefficients that solve eq. (\ref{crosscap}) to the two basic matrices
$S$ and $P$ underlying the construction, so that
\be
\Gamma_a \ = \ {{P_{1a}} \over {\sqrt{S_{1a}}}} \qquad , \label{gamma}
\ee
and display the occurrence of the new integral-valued tensor
\be
Y_{abc} \ = \ \sum_{d} \ {{S_{ad} \ P_{bd} \ {P^\dagger}_{cd}} \over {S_{1d}}}
\label{ymat}
\ee
in the Klein-bottle and M\"obius amplitudes.  Amusingly, $Y_{abc}$ yields
an additional representation of the fusion algebra.
For the $A$ and $E_6$ models,
one may also choose the Klein-bottle projection of eq. (\ref{kleinac}).  This
alters eqs. (\ref{kclass}) - (\ref{mtclass}), and the
corresponding crosscap coefficients are then
\be
{{\Gamma}^{\prime}}_a \ = \ { (-1)^{a-1 \over 2}
{P_{1,k+2-a}} \over {\sqrt{S_{1,k+2-a}}}}
\qquad .
\label{gammac}
\ee

In conclusion, we have shown that the setting of
ref. \cite{car} can encompass the construction of open descendants
for all $SU(2)$ WZW models.  In particular, we have described
in some detail how open descendants may be associated to the
$A$, $D_{even}$ and $E_{even}$ models. The resulting open spectra follow
the familiar pattern based on the fusion algebra \cite{bs}, but
whenever the closed spectrum contains
half-odd-integer isospins a totally symmetric
Klein-bottle projection leads to additional models where
pairs of real Chan-Paton charges are replaced with pairs of
complex ones. The other models, $E_7$ and $D_{odd}$, also admit
open descendants, but of a more unconventional nature, and
will be described elsewhere \cite{pss}.
\vskip 24pt
\begin{flushleft}
{\large \bf Acknowledgements}
\end{flushleft}

G.P. and A.S. are grateful to the ``Centre de Physique Th\'eorique'' of the
\'Ecole Polytechnique for the kind hospitality extended to them while this
work was being completed. This work was supported in part by
E.E.C. Grant CHRX-CT93-0340.

\vfill\eject

\end{document}